\documentclass[twocolumn,preprintnumbers,amsmath,amssymb]{revtex4}
\usepackage{graphicx}
\usepackage{upgreek}
\usepackage{color}
\usepackage{xcolor}
\usepackage{comment}
\usepackage[colorlinks=false, linkbordercolor=red, citebordercolor=green, urlbordercolor=cyan, pdfborderstyle={/S/U/W 1}]{hyperref}

\begin{document}

\title{Nonclassical light from large ensembles of trapped ions}
\author{P. Ob\v{s}il$^{1}$, L. Lachman$^1$, T. Pham$^{2}$, A. Le\v{s}und\'{a}k$^{2}$, V. Hucl$^{2}$, M. \v{C}\'{i}\v{z}ek$^{2}$, J. Hrabina$^{2}$, O. \v{C}\'{i}p$^{2}$, L. Slodi\v{c}ka$^{1}$}\email{slodicka@optics.upol.cz} \author{R. Filip$^1$}

\affiliation{
$^1$ Department of Optics, Palack\'{y} University, 17. listopadu 12, 771 46 Olomouc, Czech Republic \\
$^2$ Institute of Scientific Instruments of the Czech Academy of
Sciences, Kr\'{a}lovopolsk\'{a} 147, 612 64 Brno, Czech Republic}

\date{\today}

\begin{abstract}

The vast majority of physical objects we are dealing with are
almost exclusively made of atoms. Due to their discrete level
structure, single atoms have proved to be emitters of light
which is incompatible with the classical description of
electromagnetic waves. We demonstrate this incompatibility for
atomic fluorescence when scaling up the size of the source ensemble,
which consists of trapped atomic ions, by several orders of magnitude.
The presented measurements of nonclassical statistics on light
unconditionally emitted from ensembles containing up to more than
a thousand ions promise further scalability to much larger emitter
numbers. The methodology can be applied to a broad range of
experimental platforms focusing on the bare nonclassical character
of single isolated emitters.

\end{abstract}

\maketitle

In 1905 Albert Einstein discovered that light can be understood as
a stream of interfering photons~\cite{einstein1905erzeugung}.
At that time it was a contradiction to Maxwell's theory of light
waves~\cite{maxwell1865dynamical}. Quantum optics, however,
merges our understanding of both wave and particle aspects
together~\cite{mandel1995optical}. It seems to be commonly
considered that light produced by large number of emitters is
expected to be a mixture of classical
waves~\cite{glauber2007quantum}. This has been certified by
a number of experiments analyzing various light sources from
sun-light to laser radiation even with arbitrarily small
intensity~\cite{tan2014measuring}. On the other hand, light
from a single emitter with a transition between two discrete
energy levels of atom or solid-state object is always
nonclassical. Detection of an individual photon emitted from a single
emitter cannot be described by a mixture of classical waves and
therefore many results of single-photon experiments contradict
classical coherence theory~\cite{born1980principles}. The most
basic contradiction is observation of the indivisibility of a single photon at a beam splitter, which does not happen for classical
light waves~\cite{kimble1977photon, grangier1986experimental}.
However, the situation is already different for two photons as they can be
split by linear optics. It is evident that ideal particle
indivisibility typically manifested by measurement of perfect
anti-bunching in the case of single-photon input is lost for a large
number of photons even though some particle-like nonclassical
features may remain. On the other hand, the state of a finite
number of photons cannot be modelled as a mixture of classical
waves, as those are always based on distributions of photons up to
infinity. This particle-type of nonclassicality can remain hidden
since the limitation in number of photons can be far away from
what is detectable, as can be seen on classical sources consisting
of enormous amounts of quantum emitters.

Observability of truly particle aspects of light therefore
typically vanishes in everyday macroscopic reality and we
frequently observe only mixtures of classical waves. This is due to
the inevitable enhancement of effects which deteriorate the
macroscopic photon samples, but also affect their macroscopic
sources and detectors. However, those effects are not fundamental
and macroscopicity itself does not smudge the particle features.
In order to succeed with the detection of particle-like nonclassical
aspects of light for large number quantum emitters, the employed
detection setup and evaluation procedure must provide unambiguous
identification and sufficient information about nonclassical light
without any prior assumptions. In addition, the measured light
source needs to be stable in the number of photon emitters and the
detection efficiency of unconditionally emitted light
has to be sufficiently high to overcome the effects of background
thermal noise. The detection apparatus and the employed criterion
should be able to detect nonclassicality for a large number of
photons in the presence of high loss, finite amplitude of added
thermal noise and within feasible measurement times. These
requirements have long forced the particle-like nonclassicality of
radiation from large ensembles of emitters and its possible
applications to remain in the domain of theoretical
considerations. Apart from a few notable recent stimuli, which provide
substantial evidence that nonclassicality is not necessarily bound
to a few emitters~\cite{shcherbina2014photon,rogers2013multiple}, experimental
observations have been concerned by the analysis and control of
nonclassical features on very small number of emitters. The
ingenious fundamental questions considering the limits on the size
of the particle samples and their number statistics, applicability
of possible nonclassicality in large ensembles for detection of
related fundamental effects like quantum phase transitions,
particle entanglement, or its possible utilization for mesoscopic
quantum computation still remain to be explored.

In this letter we present the experimental observation of
nonclassical statistical properties of light emitted from large
ensembles of single-photon emitters. We employ trapped ion
crystals as a scalable source satisfying conditions for
nonclassicality observability for large numbers of emitters. The
recently proposed exactly measurable nonclassicality criterion,
adjustable to the applied detection
scheme~\cite{lachman2016nonclassical}, is used for witnessing
the nonclassical character of emitted fluorescence. We measure the
nonclassicality witness on emitted light for both pulsed and
continuous laser excitation and verify the value of
the nonclassicality threshold by conducting the same measurements with
laser light.

\begin{figure}
\begin{center}
\includegraphics[width=87mm]{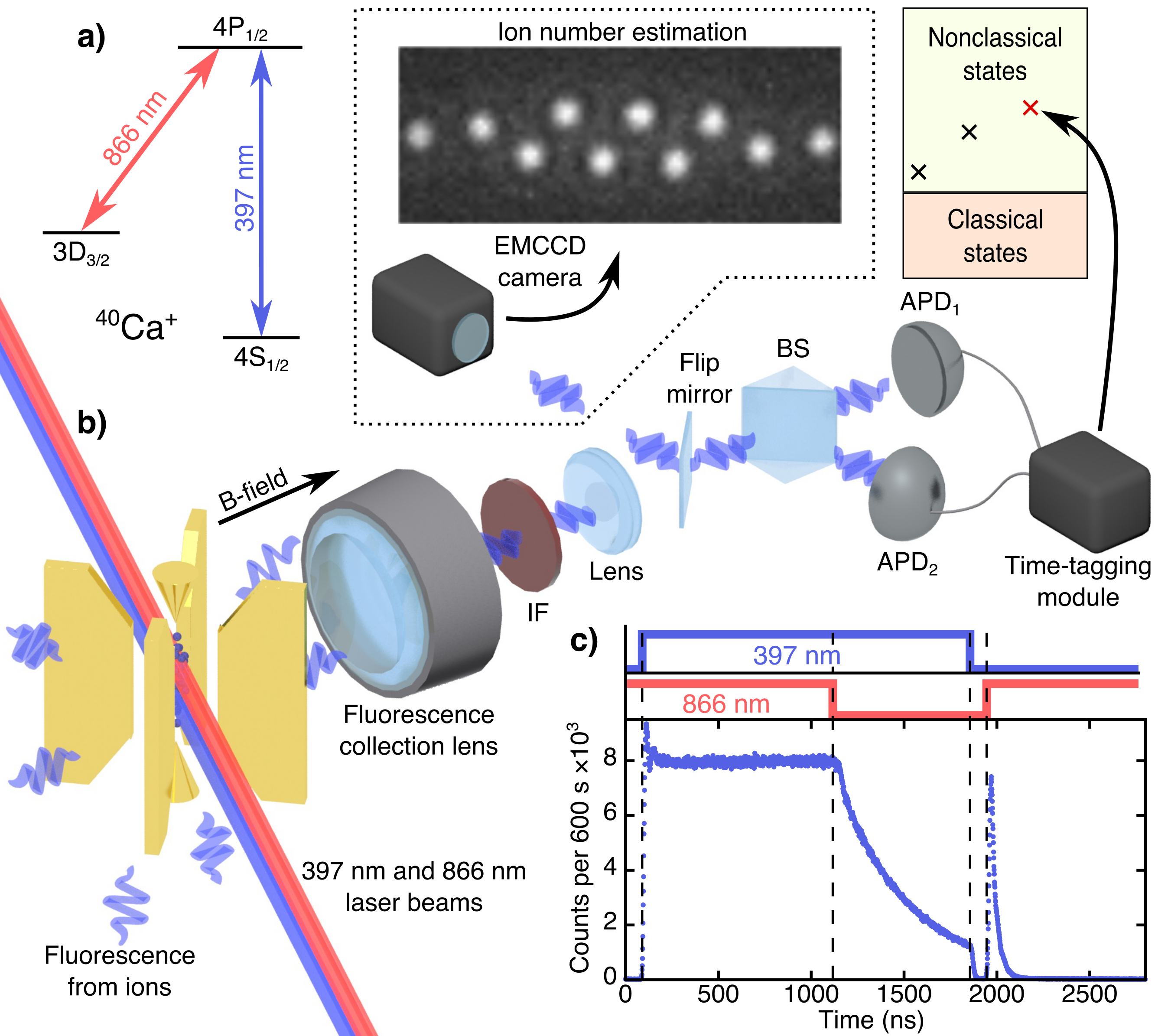}
\caption{The simplified scheme of experimental setup and laser
pulse sequence for generation and detection of nonclassical light.
a) The energy level scheme of the $^{40}$Ca$^+$ ion including
employed transitions with their respective wavelengths.
b) An ensemble of ions is trapped in the linear Paul trap and the
397\,nm fluorescence emitted along the applied magnetic field
direction (B-field) is collected by a lens objective and separated
from the 866\,nm light by an optical interference filter (IF). The
fluorescence is then directed towards the nonclassicality analyzing
setup comprising a single beam-splitter (BS), a pair of avalanche
photodiodes (APDs) and the time-tagging module. The trapped ion
crystals can be imaged on an EMCCD camera for the purposes of ion
number estimation and optimization of trapping stability. c) The
generation of analyzed light from the trapped ion ensemble in the pulsed
regime begins by Doppler cooling period, in which both lasers are
switched on. It is followed by the optical pumping stage, where
the populations are shuffled to the metastable 3D$_{3/2}$ state using only the 397\,nm laser. The analyzed fluorescence at
the 4P$_{1/2}\leftrightarrow$4S$_{1/2}$ transition is then generated
by fast depopulation of the 3D$_{3/2}$ state using the 866\,nm
laser pulse.} \label{fig:block_scheme}
\end{center}
\end{figure}

Ensembles of large numbers of ions trapped in a Paul trap possess
crucial advantages for initial proof-of-principle tests of
nonclassicality~\cite{lachman2016nonclassical, filip2013hierarchy}. First, average
trapping lifetimes of single ions in our setup exceed
several days which allows for direct observation of emitted
fluorescence even for very large ion
numbers~\cite{hasse1991structure, drewsen1998large, totsuji2002competition, okada2010characterization}. Second, laser cooled and
trapped ion crystals can constitute isotopically pure samples of
ions with lambda-type electronic level schemes. Due to their superb
isolation from environment and long trapping lifetimes, these
systems have been employed for some of the pioneering tests of
single-atom fluorescence
nonclassicality~\cite{diedrich1987nonclassical,schubert1992photon}
and more recently, led to the demonstration of single-photon sources
with record-breaking single-photon
content~\cite{higginbottom2016pure,maunz2007quantum,kurz2013high,barros2009deterministic}.
Furthermore, the typical inter-atomic distance of ions in the
traps is limited to be much larger than the wavelengths of the
involved optical transitions due to the Coulomb repulsion, which
suggests, that spontaneous build-up of collective effects can be
neglected and ions can be treated as mutually independent
emitters~\cite{brewer1995two}. In addition, the ion trapping
apparatus allows for near perfect control of the number of emitters in
the crystal due to easily repeatable, albeit probabilistic ion-loading procedures.

In our measurements we focus on reaching the maximal number of
participating ions while still unambiguously demonstrating the
nonclassicality of the emitted light field allowed by practical
limits related to the ion trapping and light collection apparatus.
The simplified scheme of our experimental setup is shown in
Fig.~\ref{fig:block_scheme}. The $^{40}$Ca$^+$ ion crystal is
created by application of the trapping and Doppler-cooling forces
in the potential minimum of a linear Paul trap. The light
scattered from ions is collected using a lens covering $\approx$2~\%
of the full solid angle with the radial position and focal point
carefully optimized for maximizing the fluorescence detection
efficiency from a single ion trapped in the trap center. See
Supplemental Material, Sec.~A 
for more experimental details.


Nonclassical features of atomic fluorescence coming from trapped
ion crystals are estimated by statistical analysis of the recorded
time-tagged detection signal corresponding to exact times of
photon arrival at the two avalanche photodiodes (APDs) using the
criteria from Lachman et al.~\cite{lachman2016nonclassical}
based on the bare estimation of the true photon detection
probabilities. These criteria fundamentally differ from
the measures that incorporate moments of photon distribution
including commonly employed nonclassicality estimation
methods based on measurements of the intensity correlation function
$g^2(\tau)$. The estimation of $g^2(\tau)$ corresponds to the
measurement of the photon number variance, which cannot be safely
realized when using binary single-photon detectors for observation of small nonclassicality from the large number of emitters. It generally
requires the estimation of the whole photon number distribution.
Although this is usually approximated by the probability of click and
double click in the limit of small photon flux, such
simplification is not safe in general and might become misleading
especially when the number of emitters, and thus, number of emitted photons increases substantially~\cite{short1983observation,
avenhaus2010accessing}.

The employed criterion~\cite{lachman2016nonclassical} includes
the real response of single-photon detectors and takes into account
possible unequal quantum detection efficiencies. It is
operationally derived from first principles by construction of
a linear functional depending on the probability of detecting no
photon on both detectors, denoted $P_{00}$, and no photon on one
particular detector, denoted $P_0$. The linear functional has the
form
\begin{equation}
F_a(\rho)=P_0+a P_{00},
\end{equation}
where $a$ is a free parameter. If we consider a symmetrical
detection scheme, optimizing $F_a(\rho)$ over all classical states
$\rho=\int P(\alpha)\vert \alpha \rangle \langle \alpha \vert
\mathrm{d}^2\alpha$, where $P(\alpha)$ is a density probability
function, leads to the threshold function $F(a)=-1/(4a)$ which covers
all classical states. The nonclassicality condition requires $a$
such that the detected probabilities satisfy $P_0+a P_{00}>F(a)$.
This $a$ can be found if and only if $P_0 - \sqrt{P_{00}} > 0$. It
thus allows for an unambiguous test of nonclassicality of light even
with a high mean number of photons, where the approximation of
moments by probabilities of clicks is not generally safe, because it can potentially imitate the nonclassical behaviour~\cite{sperling2012true}. The nonclassicality is then
witnessed by estimation of the probabilities within given time bin
period and evaluation of the distance from the nonclassicality
threshold
\begin{equation}
d = P_0 - \sqrt{P_{00}} > 0. \label{eq:nonclassical_distance}
\end{equation}
We note, that the parameter \emph{d} by no means gives a
quantitative measure of nonclassicality, it is solely a suitable
witness for nonclassical states from the large ensembles of
emitters. The parameter \emph{d} can be equivalently
defined also in terms of the probability of a click $P_s$ and a double
click $P_c$, see Supplementary Information~D for more details. However, the parameter $d$ is better for understanding the experiments with many emitters. The value of $d$ increases with the number of contributing single-photon emitters while it is non-increasing if noise, Poissonian or thermal, is added instead~\cite{lachman2016nonclassical}. The $g^{2}(0)$ does not provide such information directly, as it converges to unity for addition of both single-photon emitters and noise sources.

We measure the statistics of emitted fluorescence in both pulsed
and continuous excitation regimes, which effectively represent
different sources of radiation. In the continuous case, ions in
the crystal emit fluorescence at random and mutually uncorrelated
times and, in principle, finite linewidth of the employed
transition and laser light leakage can result in multiple
emissions from a single ion within the same time bin. This is well
illustrated by the decreased purity of the single photons emitted from single atoms
estimated from measurements of intensity correlation
functions $g^2(0)$ in continuous schemes
compared to pulsed sources, in which the multiphoton emission is
typically prohibited by the optical pumping mechanism
\cite{higginbottom2016pure,maunz2007quantum,kurz2013high,barros2009deterministic}.
The pulsed driving is thus much more convenient for demonstration
of the nonclassical emission from large atomic ensembles, provided
that the rate of the photon emission given by pulse sequence
length can be kept comparably high. As can be seen in
Fig.~\ref{fig:block_scheme}-a) and~\ref{fig:block_scheme}-c), we
employ an effectively three-level energy structure of the
$^{40}$Ca$^+$ to minimize the multiphoton content in the given
measurement time-bin by optically pumping the atomic population to
the metastable 3D$_{3/2}$ level from where the photon
emission is initiated by the 866\,nm laser pulse. The optical pumping
characteristic time is relatively long for our excitation
parameters and crystal spatial extensions, and depending on the
laser settings, about one fifth of emitters remain in the
4S$_{1/2}$ level.
The efficiency of the optical pumping to the 3D$_{3/2}$ manifold is estimated from the observed fluorescence rates in the pulsed photon generation sequence, see the example in the Fig.~\ref{fig:block_scheme}-c). The residual average detected photon rate at the end of the 397~nm pulse is compared with the value of the photon rate during the Doppler cooling period, which gives a lower bound on the 3D$_{3/2}$ population. The exact value can be then estimated by evaluating the steady state populations of the 3D$_{3/2}$ state for given Doppler cooling laser excitation parameters.
The single-photon detectors are gated with the
gating time optimized to comprise most of the generated 397\,nm
light. The detailed description of measured data processing can be
found in Supplemental Material, Sec.~B. 

\begin{figure}[!t!]
\begin{center}
\includegraphics[width=85mm]{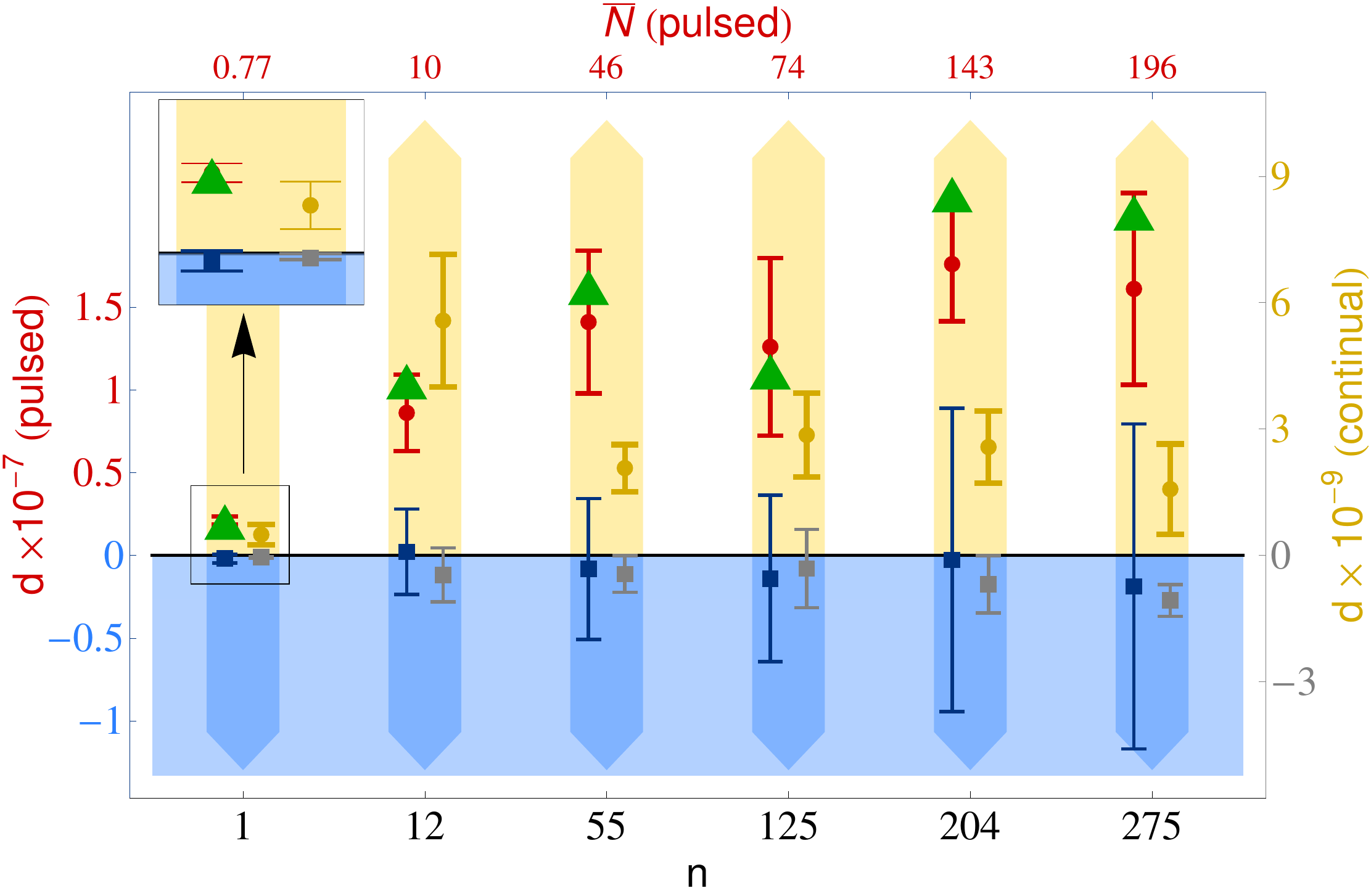}
\caption{The measured violation of nonclassicality as a function
of the number of trapped single-photon emitters and the same
measurements realized just with the laser light. All measured
values of the witness $d$ for trapped ion ensembles in pulsed (red
circles) and continuous (yellow circles) regimes are provably in
the nonclassical region given by $d>0$. The green triangles
correspond to the numerical simulation of the witness value in the
pulsed regime, details of the simulation can be
found in Supplemental Material, Sec.~C. Estimated mean photon numbers per single emission pulse $\bar{N}$ in the pulsed
regime are shown on the top axis. \emph{d} for the 397\,nm laser
light scattered from the trap electrodes has been measured using
the same detection scheme in both excitation regimes. The laser
intensity has been set to reach the detection count-rates
corresponding to measurement on light from a given trapped ion
ensemble. The dark blue and grey squares are measurements on the
laser light in the pulsed and continuous regimes, respectively.
All measurements corresponding to a given ion number are grouped
in arrow-like shaded regions. Error bars correspond to one
standard deviation.} \label{fig:nonclassicality_graph}
\end{center}
\end{figure}

The evaluated value of witness \emph{d} defined in Eq.~(\ref{eq:nonclassical_distance}) for trapped ion crystals
containing from one to up to several hundreds of ions is plotted
in Fig.~\ref{fig:nonclassicality_graph}-a). A clear violation of
the nonclassicality condition by several error bars marking single
standard deviation is observed for all ion numbers up to 275 in
both continuous and pulsed regimes. The corresponding theoretical
prediction of the parameter $d$ in the pulsed regime has been
evaluated and plotted by taking into account the measured
distribution of detection efficiencies for ions in the given
crystal. The theoretical prediction agrees well with the measured
data, even without any free fitting parameter and without taking
into account experimental imperfections like the optical pumping
spatial distribution, or intensity fluctuations of the exciting
lasers. In the presented simulation, we have considered the ion
crystals with concentric shell structure supported by our
observations. Details of the simulation can be
found in Supplemental Material, Sec.~C. 
The ion storage lifetime and crystal stability can play a crucial
role for the statistical properties of the emitted light. We have
analyzed the rate of loss of ions in our setup in the regime where
the cooling lasers frequencies are locked and the trap input
radio-frequency power remains stable. The number of trapped ions was precisely counted before and after each photon-counting
measurements and there has been no observable loss of ions in the
presented measurements. We note, that only 2 out of 14 measurement
runs were discarded due to loss of ions caused by cooling laser
frequency instability. For comparison, the quality of our single-photon
emitter-single ion has also been evaluated using the conventional measure based on the estimation
of the intensity correlation function $g^2(0)$, which gives
$g^2(0)=0.081$ in the continuous case for a 1\,ns time bin and
$g^2(0)=0.032$ in the pulsed excitation regime, comparable to other
realizations of single-photon sources with single trapped
ions~\cite{higginbottom2016pure,maunz2007quantum,kurz2013high,barros2009deterministic}.

The internal atomic dynamics and multi-photon contributions make
it difficult to theoretically predict the measured parameter
\emph{d} in continuous regime. The actual uncertainty in the
emission time and possibility of multiphoton contributions from
single atoms within finite time bins make this regime closer to
a large class of optical sources with uncontrolled internal dynamics
or partial coupling to environment~\cite{predojevic2014efficiency} and
the presented measurements will likely stimulate investigations of
their statistical properties.

The technical limit on further increasing the number of emitters in
our experiment is given by the effective detection volume of the
employed optical detection setup. The photon detection efficiency
falls rapidly for ions positioned in the radial direction from the
collection lens focal point and less rapidly, but still
considerably, along the lens symmetry axis. See Supplemental Material, Sec.~A 
for more details. The limiting
radial size of the detection volume for our detection arrangement
is 4.6\,$\upmu$m (FWHM). The detection efficiency variation at
various distances from the lens focal point suggests that increasing
the measured crystal size beyond hundreds of ions in our setup
inevitably places some of them into regions with extremely small
relative collection efficiency. The transition to small relative
detection efficiencies is smooth in all three spatial directions,
which brings in a technical question of how many emitters actually
substantially contribute to the detected photon flux. This fuzziness
in the number of contributing emitters is seen as a sign
of dealing with a system which approaches the fragile borderline
between the applicability of quantum and classical descriptions,
at which the suitability of the discrete quantum description of
some important physical variables, like number of emitters or
total energy, naturally deteriorates. The mere technical limit of
the employed optical detection setup can be eliminated by the use
of optimized imaging configurations. We demonstrate further
scalability of the nonclassicality measurements in similar setups
by changing the lens configuration so that we decrease its overall
magnification factor, which corresponds to an increase of the
radial detection volume. We reach a radial field of view of approximately
$20\,\upmu$m without observing any substantial change of the
absolute detection efficiency of an ion positioned on the optical
axis. With this configuration, we have measured the positive
distance $d=(9.48 \pm 3.93)\times 10^{-7}$ for 1500$\pm 200$ ions
in an equal 5 hour long experimental run. We note, that this would correspond to 391 ions when considering only emitters contributing to the detected optical signal with relative efficiency higher than $\eta_{\rm max}/e$, where $\eta_{\rm max}$ is the overall detection efficiency for an ion positioned at the optical axis of the employed detection system and $e$ is Euler's number.

The photon flux per single emission pulse at the input of the detection apparatus has been
estimated for each measurement in the pulsed regime from the
number of trapped ions $n$ and measured finite efficiency of the
optical pumping~$\eta_p$ as $\bar{N}=n\times\eta_p$. The highest
mean photon number at the input of our detection apparatus is
$\bar{N}=196$ photons, which correspond to $n=275$ ions and
$\eta_p=71$\,\% optical pumping efficiency. To the best of our
knowledge, this corresponds both to the largest ensemble of single-photon emitters and the largest photonic field for which nonclassical
statistical properties have been demonstrated. Furthermore, as can
be seen in Fig.~\ref{fig:nonclassicality_graph}-a), the relative
uncertainty of the measurements of the distance \emph{d} scales
very favorably with number of ions, which promises scalability of
our measurements to much higher ion numbers, provided that the
emitted light is collected efficiently.

Furthermore, there seems to be no fundamental limit on further
substantial increase of any of these quantities in a similar
experimental apparatus. We have simulated the emission of light
from a crystal containing up to $10^5$ ions. The simulation predicts
scaling of the parameter $d$ and uncertainty caused by finite
measurement with size of the crystal. As already predicted in
Ref.~\cite{lachman2016nonclassical}, the main limiting
parameter for unambiguous detection of nonclassicality for stable
ensembles of single-photon emitters is the measurement time. The
detailed analysis presented in Supplemental Material, Sec.~C 
shows, that the required
experimental time does not grow substantially until the mean
photon flux at photo-detectors reaches several photons. The
nonclassicality of stronger light would be also observable,
however it would require additional attenuation to reach the
optimal photon flux for measuring nonclassical properties.

The presented measurements demonstrate the first unambiguous proof of the
nonclassical character of light fields emitted from a large ensemble
of single-photon emitters. We have shown that ensembles consisting
of emitters which individually produce nonclassical light, keep
this statistical property when scaling up their size by at least
two orders of magnitude by trapping and measuring for a range of
ion numbers, from a single ion up to 275 ions. Moreover, the demonstrated
nonclassicality measurements present robustness against many
imperfections of individual sources. We have also verified several
aspects of emission from ensemble of single-photon sources
compared to emission from a single one. Most notably the
measured nonclassicality witness value in the pulsed regime grows
or stays approximately constant with increasing emitter
number.

Our experimental test opens the possibility of searching for
nonclassical light emission from recently developed ensembles of
atomic~\cite{northup2014quantum,neuzner2016interference} and
solid-state
emitters~\cite{shcherbina2014photon,rogers2013multiple} and
studying their internal dynamics from a new
perspective~\cite{bhatti2017superbunching,jahnke2016giant,hornekaer2002formation}.
It will allow their exploration before single emitters are
isolated and further direct exploitation in quantum
technology~\cite{palacios2017large}. Furthermore, due to the large
dependence of the ability to detect nonclassicality on the
statistics of emitters and emitted light mode-structure stability,
the presented scheme can be easily applied for detection of phase
transitions from solid to gas or plasma phase~\cite{hornekaer2002formation}. The presented
demonstration can be directly extended in the future to ion numbers
beyond thousands by employing optimized optical fluorescence
collection schemes, and later, to observations of emissions going
toward the Fock states of indistinguishable photons from atoms
inside a cavity~\cite{casabone2015enhanced,sayrin2011real}. It
substantially shifts the range of energies in which observable
manifestations of discrete quantum features of light and matter
should be anticipated and thus will likely trigger the
construction of truly macroscopic and intense sources of quantum
light.

In the course of preparing our manuscript we became aware of
another experiment studying nonclassicality from ensembles of
single-photon emitters~\cite{moreva2017direct}. It demonstrates the applicability of the presented proof-of-principle methodology for studies of emission from clusters of NV centers in diamond and its robustness against realistic sources of noise.

\begin{acknowledgments}
The work reported here has been supported by the grant No.~GB14-36681G of the Czech Science Foundation. We acknowledge the kind technological support from the group of Rainer Blatt.
\end{acknowledgments}


%

\clearpage

\newpage

\setcounter{equation}{0}
\setcounter{figure}{0}
\setcounter{table}{0}
\setcounter{page}{1} \makeatletter
\renewcommand{\theequation}{S\arabic{equation}}
\renewcommand{\thefigure}{S\arabic{figure}}
\renewcommand{\theHequation}{Supplement.\theequation}
\renewcommand{\theHfigure}{Supplement.\thefigure}
\renewcommand{\bibnumfmt}[1]{[S#1]}
\renewcommand{\citenumfont}[1]{S#1}

\section*{Supplementary information: Nonclassical light from large ensembles of trapped ions}

\subsection{Experimental setup}    
\label{subsec:exp_par}

{\it Ion trapping} \\

The trap parameters are set with respect to maximizing the
stability of a  particular ion crystal and the number of detected photon
counts. This corresponds to close to symmetrical configuration of
oscillation frequencies of the trapping potential $\omega_x
\approx \omega_y \approx \omega_z \approx 778$ kHz for measured
ion crystals containing more than 12 ions, and configuration with
$\omega_x \approx \omega_y \approx 1341$\,kHz and $\omega_z
\approx 778$\,kHz for measurements with 1 and 12 ions. Here,
$\omega_z$ is the frequency of the trapping potential in the axial
direction of the trap, that is, along the axis connecting the two
tip electrodes, and the $\omega_x, \omega_y$ are the frequencies
of the potential in the radial directions which lie in the plane
perpendicular to the axial direction. The measured ion ensembles
with more than 12~ions correspond to near spherical crystals with
concentric shell structure~\cite{drewsen1998large_s,hornekaer2002formation_s,okada2010characterization_s},
which have been chosen to optimize the overall scattered light
collection efficiency in the direction perpendicular to the trap
axis. The ions are Doppler cooled by a red detuned 397\,nm laser
exciting the 4S$_{1/2}$ $\leftrightarrow$ 4P$_{1/2}$ transition
and a 866 nm laser beam is simultaneously used for repumping the
population from the metastable 3D$_{3/2}$ state to the cooling
transition. A magnetic field of 12 Gauss is applied along the
observation direction to lift the degeneracy of the 3D$_{3/2}$
state manifold and thus enable its effective depopulation.
The excitation beams are co-propagating in the direction perpendicular to the magnetic field direction with linear polarization parallel to it. \\

{\it Light detection} \\

\begin{figure}
\begin{center}
\includegraphics[width=85mm]{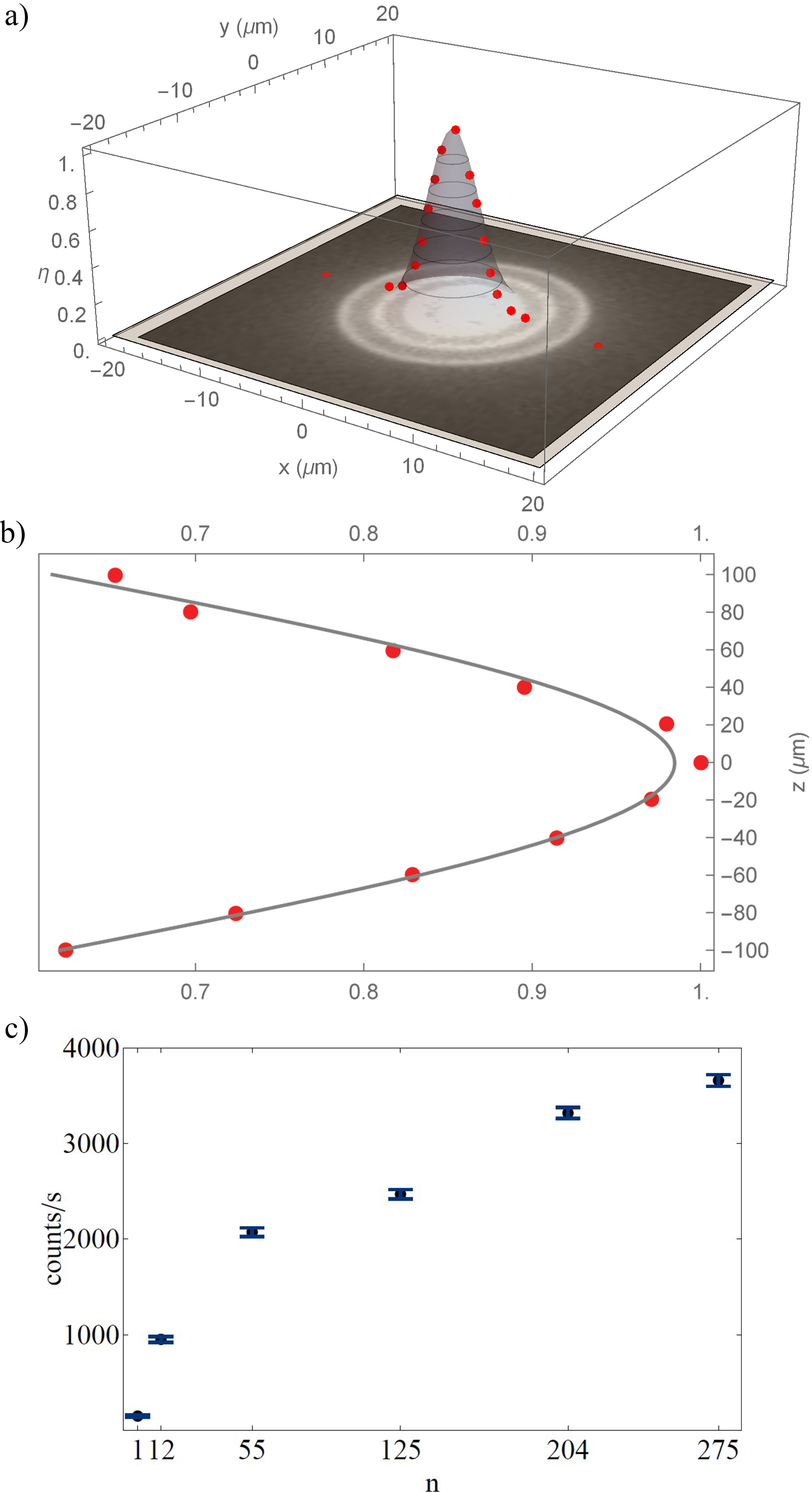}
\caption{The estimation of the effect of finite detection volume by the
measurement of detection efficiencies in the radial and axial
object planes and evaluation of the overall detected count rate
for trapped ion ensembles. In a) we show relative detection
efficiencies normalized to the maximal efficiency corresponding to
the lens on-axis position. The bottom plane of the 3D figure is
filled with the picture of the ion crystal consisting of $n=275$
ions scaled to correct units. The graph in b) shows measured
efficiencies for the ions effectively displaced along the lens
axis for the displacements relevant for the spatial sizes of the
measured ion crystals. The fitted Gaussian function has FWHM of
238 microns. The dependence of the detected count rate on the number of
ions \emph{n} shown in part c) in the pulsed regime demonstrates the
effect of decreased overall detection efficiency of light emitted
far from the collection objective optical axis.} \label{detVol}
\end{center}
\end{figure}

The light scattered from an ensemble of ions is collected by an
objective covering $\sim$2~\% of the full solid angle and
frequency filtered (IF) to isolate the 397\,nm fluorescence. The
photons are then directed towards one of two possible detection
arrangements. The EMCCD camera (Luca-S, Andor) is used for spatial
analysis of the trapped ion crystal, estimation of the number of
emitters before and after each experimental run, and further
calibration measurements, including estimation of the excess
micromotion of off-axis positioned ions and photon emission
probability homogeneity across the crystal. The light is further
split by a beamsplitter (BS) and directed towards a pair of
free-space-coupled avalanche photodiodes (APDs, Count-series,
Laser Components). The overall detection efficiencies of
fluorescence emitted by the ion positioned in the focus of the
collection lens have been estimated to be 0.033\% and 0.028\%, for
the APD$_1$ and APD$_2$, respectively. Precise arrival times of
incoming photons are recorded (PicoHarp300, PicoQuant) with time
resolution of 4 ps and further processed to obtain the individual
photon detection probabilities. The measured detection
efficiencies fall down rapidly for ions displaced radially from
the collection objective focal point due to the limited field of
view of the whole detection setup, which gives the limit on the number
of effectively coupled ions. We have separately estimated
detection efficiencies for ions displaced in radial and axial
directions. In the radial direction of the lens object plane, which
corresponds to ion displacement along the trap axis, the measured
detection efficiencies of single photons scattered from a single
ion normalized to the efficiency in the center yield an approximately
Gaussian efficiency profile with a full width at half maximum (FWHM)
of 4.6\,$\upmu$m, see Fig.~\ref{detVol}-a). This relatively narrow radial
field of view of the employed detection setup is given by the
magnification of our detection optics and size of the active
detector area. The effective image area is given by the
magnification and optical abberations of the employed optical
detection setup consisting of the high-NA objective with object
focal distance of approximately 66\,mm and an additional 400\,mm
plano-convex lens together with the finite active detection area
of the APD of about 100\,microns. In the axial direction, the
effective ion displacement has been emulated by the collection
objective axial displacement using precise translation stage. The
measured decrease of the relative detection efficiency is plotted
in Fig.~\ref{detVol}-b). The FWHM of the fitted Gaussian function is
238\,microns. As can be seen in Fig.~\ref{detVol}-c), this detection limitation manifests as small relative contributions of ions positioned far from the detection optical axis. \\

{\it Pulsed and continuous modes of excitation}\\

Measurements of nonclassicality of light emitted from ion crystals
are realized in both continuous and pulsed regimes, which
substantially differ in properties of emitted light and in the
technical feasibility of its detection with single photon
detectors with finite detection bandwidth. We have evaluated this
difference by comparison of the $g^2(0)=0.081$ in the continuous
case for a 1\,ns time bin with $g^2(0)=0.032$ in pulsed excitation.\\ 

{\it General remarks}\\

The particular excitation parameters in the continuous regime give us
at maximum up to 36900 counts/s and 31500 counts/s on APD$_1$ and
APD$_2$ detectors, respectively, far from the detector saturation
rates, which are specified to be approximately in the $10^6$\,counts/s
regime.

\subsection{Data processing}    

{\it Measurement and evaluation of the d parameter}\\

\label{subsec:data_proc} The fluorescence from the ion ensemble is
detected using two APDs and the arrival times of scattered photons are
recorded using time-tagged device with 4\,ps resolution. Further
processing has two parts, manipulation with raw time-tagged data
for estimation of probabilities $P_{00}$ and $P_0$, and estimation
of nonclassicality with corresponding uncertainties. In the first
step, we start by evaluation of the total number $N_{\rm TB}$ of
all time-bins of size $\tau$ in a given measurement, $N_C$ number of
events where both APDs register a photon in the same time-bin and
$N_{S1}$ and $N_{S2}$, which correspond to events where APD$_1$
and APD$_2$ register a photon, respectively. While in the
continuous regime we need to cut the time arrival data to
time-bins, in the pulsed measurements processing is one step
easier, because each sequence includes the APD gating pulse which
corresponds to a single time-bin. We further shorten this time-bin
in the data processing step in order to remove the false clicks
due to the APD gating pulses, which otherwise correspond to about
10\,\% of real photon detection events depending on the mean input
photon flux. According to the employed criterion of
nonclassicality, we need to estimate the probabilities $P_{00}$
and $P_0$. The measured photon detection numbers are related to
these probabilities through the following relationships:
\begin{equation}
P_{00}=\frac{N_{\rm TB}-N_{S1}-N_{S2}-N_C}{N_{\rm TB}},
\label{measP00}
\end{equation}

\begin{equation}
P_{0k}=\frac{N_{\rm TB}-N_{Sk}-N_C}{N_{\rm TB}}
\label{measP0}
\end{equation}

and
\begin{equation}
P_0=\sqrt{P_{01} \cdot P_{02}},
\end{equation}
where the use of the geometrical mean is due to a slight imbalance
between the absolute detection efficiencies in our detection
apparatus. The use of the geometrical mean here is justified by the
following consideration. For a known imbalance $T$ of the
detection efficiencies, nonclassicality is witnessed if at least
one of the following conditions is
satisfied~\cite{filip2013hierarchy_s}
\begin{equation}
P_{01}>P_{00}^T
\end{equation}
\begin{equation}
P_{02}>P_{00}^{1-T}.
\end{equation}
Multiplying left and right sides of those inequalities excludes $T$ and leads directly to
\begin{equation}
P_{01} \cdot P_{02}>P_{00}.
\end{equation}
It is obvious, that if this relation holds, at least one of these
inequalities has to be satisfied, which
guarantees the nonclassicality.

From $P_0$ and $P_{00}$ we estimate nonclassicality of the
measured light field using the
equation~(2) 
from the main part of the
manuscript. The error bars are estimated statistically by dividing
our 5~hour long measurement into 5 parts of equal length for which
the nonclassicality has been estimated independently.\\

{\it Statistical evaluation of the measured results}\\    
\label{subsec:fit}

We have statistically evaluated the behaviour of the $d$ parameter as a function of the number of emitters in our experiment measured in the pulsed excitation regime. We have performed comparisons of individual $d$ values together with the measured error propagation with the following results. $d(1) < d(12)$ with the confidence level of 99.9\% and $d(12) < d(55)$ with the confidence level of 99\% which means that the value of $d$ clearly grows up to 55 ions. The comparisons between measurements of $d$ for number of ions $n \geq 55$ are, from point to point, not conclusive with similarly high confidence levels mainly due to relatively large error bars compared to the difference in their mean values. However, one can still give a statistical estimate on the validity of the overall rising trend by estimation of the mean gradient of the weighted linear fit and its error. This analysis confirms the observation of further increase of $d$ in the presented measurements for number of ions $n \geq 55$. The value of $d$ increases with the number of ions in the range $n \geq 55$ with a confidence level of 87\%.

\subsection{Simulations}    
\label{subsec:simulations}

{\it Estimation of the theoretical distance for our experimental setup} \\

Although the source of light consisting of trapped ions is a very
complex system with a large number of inner parameters, one can
capture the properties of the emerging light using several
simplifying assumptions about the physics behind. First, each ion
is treated as an ideal single photon emitter whose emission is
independent of the presence or state of other ions. This means
that the light is radiated independently. The cooling of the ions
causes formation of the ion crystal. The crystal in our
measurement has spherical symmetry (except the crystal with 12
ions) and a shell structure. The numbers of ions in different
shells are obtainable by simulation of molecular
dynamics~\cite{hasse1991structure_s}. The occupation of shells in our crystals is
following: a crystal with 12 ions is formed only in one shell,
a crystal with 55 ions has two shells (12+43 ions), a  crystal with 125
ions has three shells (82+35+8 ions), a crystal having 204 ions
consists of four shells (115+60+25+4 ions), and a  crystal with 275
also has four shells (150+80+36+9 ions). According to the numerical
model from~\cite{hasse1991structure_s}, the spacing among particular shells is
almost constant. Additionally, we suppose that ions within shells
have no crystal structure but rather correspond to a gaseous or
plasma state with homogeneous but random distribution in each
shell~\cite{drewsen1998large_s, hasse1991structure_s, okada2010characterization_s}. This assumption is supported
by our measurements of ion ensemble images on EMCCD camera. To
simulate this random distribution, we choose several polyhedra
carrying some symmetry whose vertices, after random rotation,
determine the position of ions on a shell. Therefore the
simulation requires a sufficient number of random rotations of each
polyhedron to obtain a homogenous distribution of vertices. The
employed polyhedra are a tetrahedron (four vertices), a cube (8
vertices), a regular icosahedron (12 vertices), a regular dodecahedron
(20 vertices) and a truncated icosahedron (60 vertices). To simulate
a shell with an arbitrary number of ions, we can combine these
polyhedra and randomly delete a few vertices to adjust the total
vertices to the desired number.

\begin{figure}[!t!]
\begin{center}
\includegraphics[width=85mm]{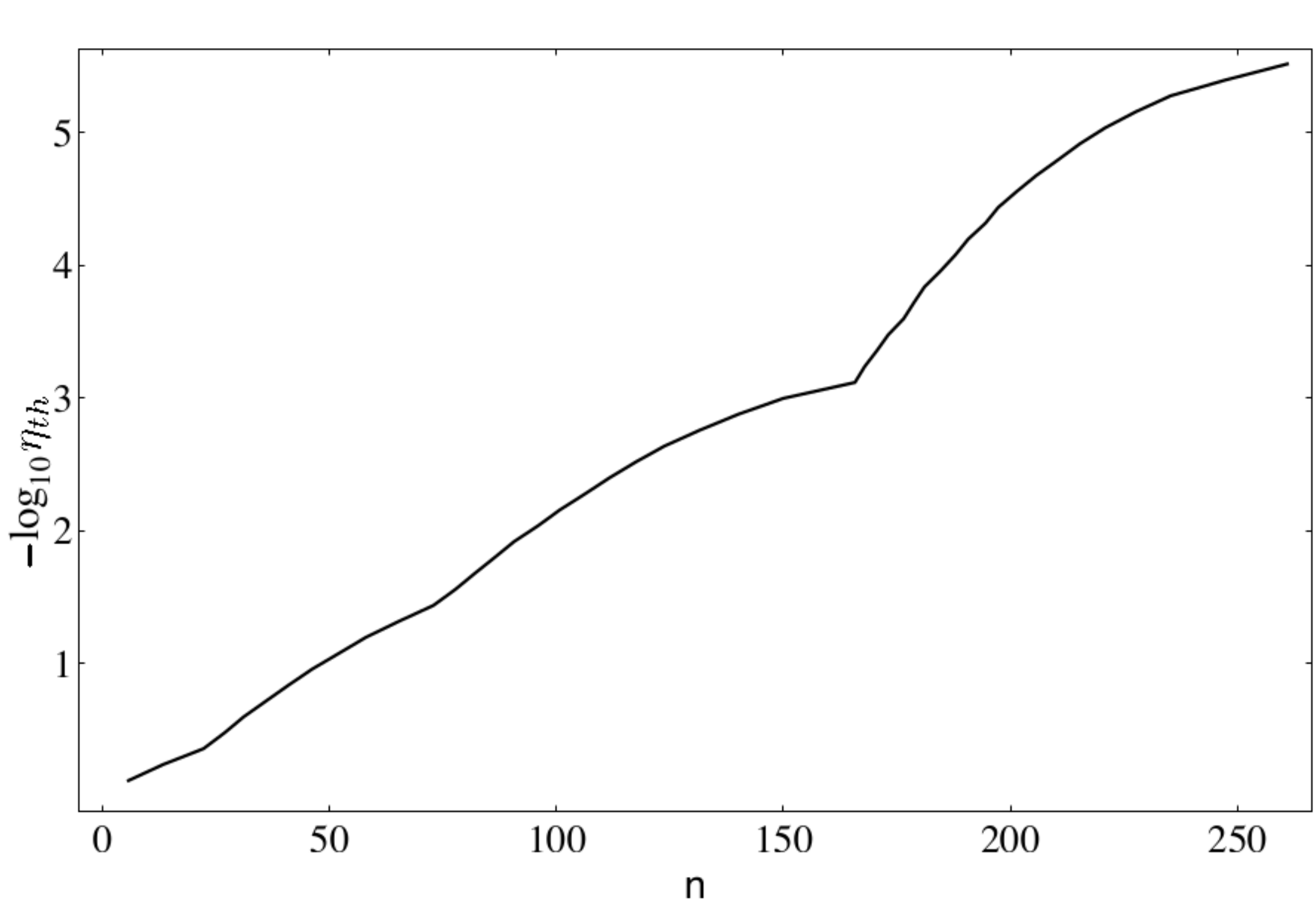}
\caption{Simulation of number of contributing emitters $n$ with
the efficiency higher than relative threshold efficiency
$\eta_{\rm th} = \eta/\eta_0$ and for the crystal
containing 275 ions.} \label{fig:photonContribution}
\end{center}
\end{figure}

If the ions are excited, they act as single photon emitters
radiating photons by spontaneous emission in directions defined by
a given dipole radiation pattern. However, the lens is able to
collect and focus on detectors only a small part of this emission.
This, together with quantum efficiency of the detectors, yields
an overall detection efficiency $\eta_0=6.1 \cdot 10^{-4}$ for the
emitters situated in the object focal point of the whole detection
optical setup. As demonstrated in measurements presented in
Supplementary information~\ref{subsec:exp_par}, emitters
positioned out of the focus are less likely to be observed due to
decreased detection efficiency. We assume that the detection
efficiency depends on position as
\begin{equation}
\eta = \eta_0 \cdot e^{-\frac{r^2}{2 \sigma_r^2}-\frac{a^2}{2 \sigma_a^2}}
\label{gauss}
\end{equation}
where $r$ is the distance from the optical axis of the lens and
\emph{a} is the distance from the focal plane. $\sigma_r$ and
$\sigma_a$ correspond to variances of this Gaussian distribution.
We have experimentally estimated their values $\sigma_r=2.3$
$\upmu$m and $\sigma_a=98$ $\upmu$m. It suffices to use only one
crystal for the calibration of the spatial size in the simulation,
because the ratio of the number of ions in a crystal to its volume is
approximately constant~\cite{hasse1991structure_s}. Significantly, this approach
assigns statistics of detector clicks to a crystal having \emph{n}
ions. The probabilities $P_{00}$ and $P_0$ employed in our criterion
of nonclassicality are obtained by
$P_{00}=\prod_{i=1}^n(1-\eta_i)$ and
$P_0=\prod_{i=1}^n(1-\eta_i/2)$. This enables us to plot the estimate
of parameter \emph{d} for crystals in our measurement as depicted
in Fig.~2 
in the main part of the
manuscript. The largest crystal in these measurements contained
$n=275$ ions. However, as described in Supplementary
information~\ref{subsec:exp_par}, not all emitters contributed
equally to the detected light. The measured detection efficiencies
as a function of radial ion displacement show that increase of
the measured crystal size beyond approximately one hundred ions
inevitably places some of them into the regions with very small
relative detection efficiency. This gives rise to questions
about the lowest detection efficiency of an ion, which is still
relevant for the observed light statistics, relative to the ion
with maximal detection efficiency. Because it cannot be answered
generally, we estimate the number of ions from which photons are
detected with probability above some value, see
Fig.~\ref{fig:photonContribution}. Apparently, the narrow volume
of the lens significantly suppresses detection of photons emitted
from outer shelves of the crystal.

{\it Prospect of measurements on higher number of trapped ions} \\

The theoretical prediction of visibility of nonclassicality for
higher photon flux requires both the estimation of the parameter
$d$ and a corresponding uncertainty that stems from the fluctuation of
experimental outputs (\ref{measP00}) and (\ref{measP0}). Let us
introduce a linear combination of the measured quantities and
their mean values
\begin{eqnarray}
X&=&2P_0-P_{00} \nonumber \\
Y&=&P_0+2P_{00} \nonumber \\
x&=&\langle X \rangle \nonumber \\
y&=&\langle Y \rangle.
\end{eqnarray}
The probability density function of random quantities $X$ and $Y$ is approximated in the limit of a high number of experimental runs by Gaussian distribution
\begin{equation}
P(X,Y)=\frac{1}{2 \pi \sqrt{V_x V_y}}\exp\left[-\frac{(X-x)^2}{2V_x}-\frac{(Y-y)^2}{2V_y}\right],
\label{densF}
\end{equation}

\begin{figure}[!t!]
\begin{center}
\includegraphics[width=85mm]{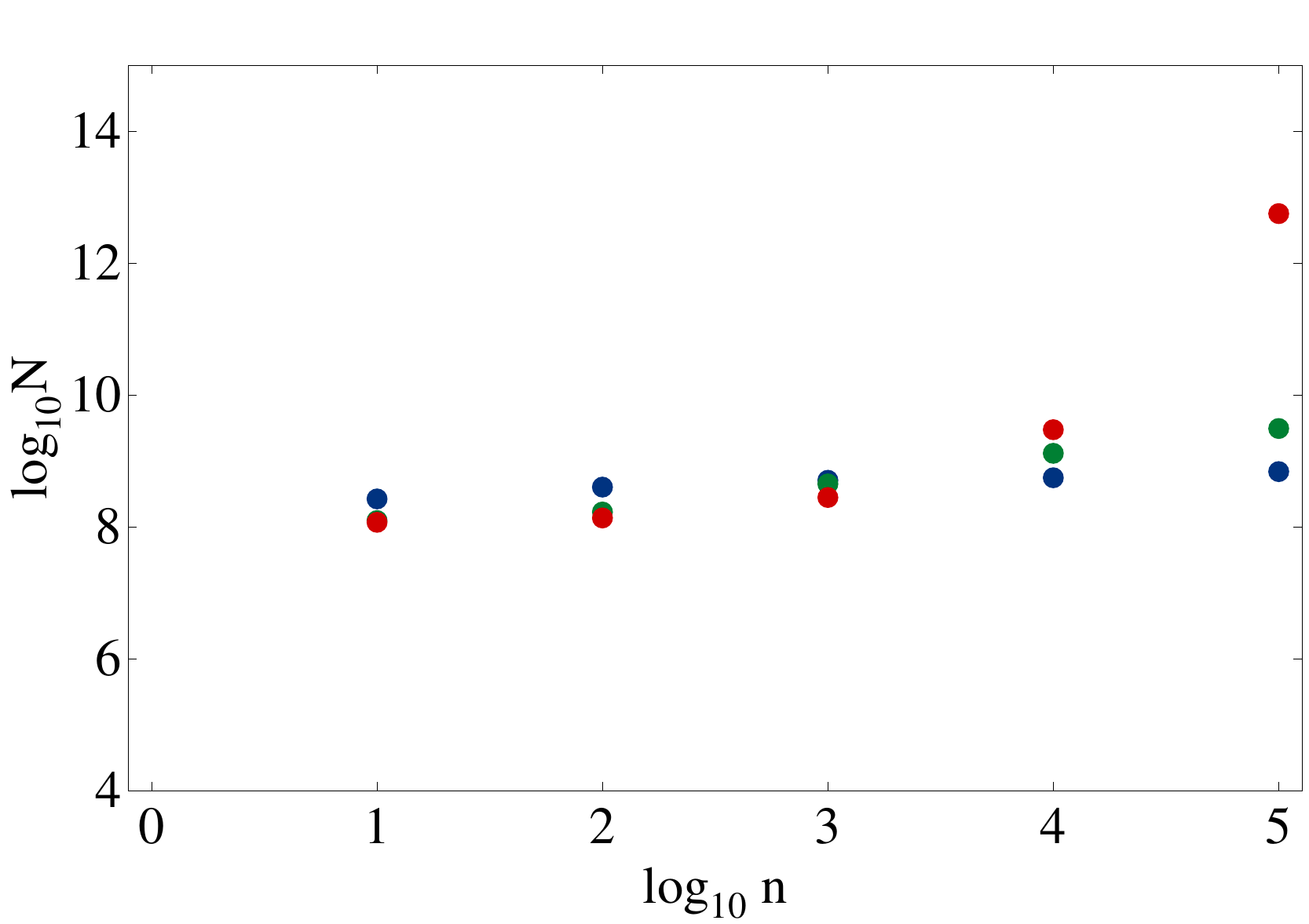}
\caption{The number $N$ of experimental runs necessary to violate
the nonclassicality by two errorbars is depicted for crystals
counting $n$ ions. The paramters of the detection are
$\eta=4\times 10^{-4}$ and $\sigma_a=98 \mu m$. Different colors
correspond to $\sigma_r=2 \mu m$ (blue), $\sigma_r=8 \mu m$
(green) and $\sigma_r=20 \mu m$ (red).} \label{fig:estimation}
\end{center}
\end{figure}

where $V_{x,y}$ is the variance of the variable $X,Y$. These
quantities are related to detection events such that
\begin{eqnarray}
\langle X \rangle &=&1-P_c \nonumber \\
\langle Y \rangle &=&\frac{1}{2} P_s+\frac{3}{2}P_{00},
\end{eqnarray}
where $P_c$ is a probability of a coincidence click and $P_s$ means probability of just a single click. Thus, the variances obey
\begin{eqnarray}
V_x&=&\frac{P_c(1-P_c)}{N}\nonumber \\
V_y&=&\frac{P_s(1-P_s)}{4 N}+\frac{9P_{00}(1-P_{00})}{4 N}.
\end{eqnarray}
The parameter $d$ is given by
\begin{equation}
d=\frac{2 X+Y}{5}-\sqrt{\frac{-X+2Y}{5}}.
\end{equation}
The probability density function (\ref{densF}) enables us to obtain
\begin{equation}
\textrm{var}(d) = V_x\left(\frac{\sin \phi}{2 \sqrt{p_{00}}}+\cos \phi \right)^2 + V_y \left(\frac{\cos \phi}{2 \sqrt{p_{00}}}-\sin \phi \right)^2,
\end{equation}
where $\phi=\arctan 1/2$. For very weak light sources, one gets
$\textrm{var}(d)\approx 0.315 P_c/N \ll \textrm{var}(P_0)$, which
indicates that the measured point fluctuates mainly in the direction
parallel with the curve corresponding to the nonclassicality
threshold.

The shell structure of a crystal simulated in \cite{hasse1991structure_s} is
presented completely for crystals containing up to 2000 ions.
Crystals with a higher number of ions tend to change the arrangement
of ions to a bcc grid \cite{totsuji2002competition_s}. The Cartesian coordinates of
emitters in the bcc grid are given by the formulas
\begin{eqnarray}
x_i &=& i u+\frac{1}{4}\left[-(-1)^j+1\right]u\nonumber \\
y_j &=& j \frac{\sqrt{3}}{2}u+\frac{1}{2\sqrt{3}}u\left[-(-1)^k+1\right]\nonumber \\
z_k &=& k \sqrt{\frac{2}{3}}u,
\end{eqnarray}
where $u$ is the distance between two ions and $i$, $j$ and $k$ are
integers. To emphasize further scalability of our measurements, we
analyze the number of experimental runs necessary for violation of
nonclassicality by two standard deviations, see Fig.~\ref{fig:estimation}.

\subsection{Nonclassicality criterion}    
\label{subsec:criterion}

The employed condition on
nonclassicality~\cite{lachman2016nonclassical_s} has been derived
ab initio without any assumptions about the state of the detected
light and can be equivalently defined in terms of the probability of
a click $P_s$ and a double click $P_c$. The condition for the
symmetric detection technique then says $P_c/P_s^2<1$. In this
symmetric case, the ratio on the left side of the inequality
converges to $g^{(2)}$ but it is never the same. In addition, the
criterion can be advantageously formulated for any asymmetry of
the detectors and can fully incorporate known structure or
properties of the employed realistic detector. Importantly, it also
allows a choice of the suitable parametrization for the analysis. The
advantage of the particular selected parametrization by
probabilities $P_0$ and $P_{00}$ is a very simple theoretical proof
that an arbitrarily large ensemble of single photon emitters
generates nonclassical light~\cite{lachman2016nonclassical_s}.


%

\end{document}